# Embodied Emissions of 2.5D and 3D Scaling in Chipletized GPUs for Low-Carbon AI

H. Watson[1], A. Singh[1], J. Stephenson[2,3], J. Soethoudt[2], J. Svedas[1], D. Bhattacharjee[2], and J. Myers[1]
[1]imec, Cambridge, UK, email: hannah.watson@imec-int.com
[2]imec, Leuven, Belgium, [3]Vrije Universiteit Brussel, Brussels, Belgium

***Abstract*—** GPU compute resources are scaling to support growing AI workloads, leading to further rises in the climate change impact of hyperscale data centers that rely on high-performance GPUs. To model how packaging decisions impact the carbon footprint we have estimated the embodied emissions (EE) of a chiplet-based GPU that has been scaled in 2.5D and 3D using the imec.**netzero** modelling platform. We doubled the number of compute units and high bandwidth memory (HBM) capacity in the 2.5D and 3D scaled systems and found that 3D scaling results in an 89% increase in EE, as compared to an increase of 141% for 2.5D scaling, positioning 3D scaling as an attractive solution to limit the embodied emissions due to GPU scaling.

## I. Introduction

Global demand for data center capacity is predicted to nearly triple by 2030, with 70% of demand associated with AI workloads. GPUs are the workhorse of hyperscale datacenters for accelerating AI workloads and are contributing a large proportion of overall emissions due to their growing design complexity and power requirements. Informed design choices can reduce the embodied emissions (EE) from manufacturing and the operational emissions related to power consumption during the use-phase. To investigate the impact of such choices for GPU scaling on EE, we have taken a chiplet-based GPU design inspired by the AMD MI300X GPU [1] as a baseline and increased the number of compute units and high bandwidth memory (HBM) per GPU in both 2.5D and 3D. Scaling compute resources in these systems can provide increased FLOP/s and memory capacity to meet the requirements of future AI workloads.

## II. Experimental Setup

EE's are estimated with imec.**netzero** [2] which uses a bottom-up approach to model a virtual high-volume IC manufacturing fab for a range of semiconductor components and technology nodes [3]. Components are defined with technology parameters, such as process node, and wafer-level packaging (WLP) flows. Current WLP flows do not include emissions for extra thinning of the DRAM dies in the HBM stacks.

Die yield for the logic components and interposer are calculated using the Murphy model with die area, process defect density (0.1 defects/cm$^2$), and critical area fraction. The critical area fraction is the proportion of the die area that is critical and vulnerable to defects. We consider a critical area fraction of 100% for the logic dies and 15% for the passive interposer. We assume a uniform defect size distribution, however, yield is likely to be higher for passive interposers as only larger defects cause failures for large feature sizes [4]. Other factors, such as reticle-stitching and warpage can also increase EE of multi-reticle sized systems, however they are beyond the scope of this work. We use fixed yields for DRAM dies and dummy Si components.

We assume that only known-good-dies are assembled into subpackages and that testing occurs when all dies are assembled. For GPU package assembly we assume that only known-good-subpackages are assembled on to the interposer before testing. Yield losses during assembly are related to connectivity failures due to TSV fabrication errors or misalignment. Failures during subpackage or GPU package assembly results in the loss of all subcomponents and subpackages. EE per component is calculated per known-good-die, where the EE for subcomponents in subpackages are adjusted according to the subpackage assembly yield, and total EE is adjusted according to the GPU package assembly yield. Compound subpackage and GPU package assembly yield are calculated assuming 99% packaging yield per die within the subpackage and subpackage attachment to the interposer.

## III. Embodied Emissions

### A. Baseline System

Our baseline system, Fig. 1, is a chiplet-based design and a mixture of 2.5D and 3D integration, contrasting with monolithic compute dies such as those used in the NVIDIA H100 GPU. Using chiplets can reduce EE because smaller dies have higher yield. We modelled the baseline GPU as: 1 passive Si interposer with 3 metal layers, 4 IO dies (IOD) with through-silicon-vias (TSVs) that are connected to the interposer with micro-bumps, 8 accelerator compute dies (XCD) placed on the IOD with hybrid bonding, and 8 micro-bumped stacks of 12-Hi HBM3 on top of 8 HBM3 logic dies. We used the following process nodes: 5 nm for the IOD, 5 nm with 4 additional metal layers for the XCD, 28 nm for the HBM3 logic dies, and DRAM1z for the DRAM dies. Dummy Si components are used for structural stability and thermal dissipation. The subpackages in this system are the IOD-XCD and HBM 3D stacks, shown in Fig. 1c. Die dimensions and process nodes are estimated from public sources. Fig. 2 shows the EE of the baseline system where the largest contribution is from the DRAM dies in the HBM stacks, 53.2%, followed by the logic dies, 28%.

### B. *Scaling Compute Resources In 2.5D*

We consider a 2.5D scaled system with a 100% larger package, shown in Fig. 3, doubling HBM capacity and number of XCD. The larger package is ~7x reticle size (33x26mm) where the interposer yield drops by 21.4% to 43.5%. Subpackage yield for HBM and IOD-XCD stacks is the same as the baseline case, but the GPU package assembly yield drops by 10% to 78.6% because the 2.5D scaled system has double the number of subpackages on the interposer as the baseline system. The EE of the 2.5D scaled system is 141% higher than the baseline system, shown in Fig. 5a.

### C. *Scaling Compute Resources In 3D*

The proposed 3D scaled system, Fig. 4a, has 8 additional XCDs on top of the existing XCDs using hybrid bonding, along with 12 stacks of 16-Hi HBM that is in line with the upcoming HBM4 specification. We increased the GPU package size by 25% (~4x reticle size) to match the number of XCD and HBM capacity of the 2.5D system. The number of compute units may vary with chip design, partitioning scheme used between the top and bottom XCD, and the area required for TSVs in the bottom XCD. Increasing interposer area by 25% results in the interposer die yield dropping 6.4% to 58.5%. We use the same DRAM process node as the baseline case and only vary the stack height. Subpackage yield of the IOD-XCD stack drops from 97% to 95.1%. Increasing HBM stack height from 12 to 16 reduces the HBM subpackage yield by 3.5% to 84.3%. This does not include yield losses from the increased risk of warpage or cracking due to increased die thinning needed for 16-Hi stacks. The 3D scaled system has 4 extra subpackages on the interposer, resulting in GPU package assembly yield dropping by 3.5% to 85.1%. We find that EE is 88.5% higher than the baseline system, shown in Fig. 5a.

### D. *2.5D vs. 3D Scaling Comparison*

The EE of the 2.5D scaled system is 28% higher than the 3D scaled system, largely due to a higher number of required IOD, a larger interposer, and more subpackages on the interposer. The IOD has the highest EE per active die because it has the highest die area. 3D stacking of XCD allows for fewer IOD-XCD subpackages and so fewer IOD. Resulting in the 2.5D scaled system having double the number of IOD as the 3D scaled system. Systems with fewer subpackages can be more compact, increasing GPU package assembly yield. Fig. 7 shows that losses in the final assembly stage increase EE for the 3D scaled system by 18%, as opposed to 27% for the 2.5D scaled system. Adding more dies in the 3D stacks reduces subpackage yield, increasing EE per subpackage, however, the total EE from subpackages is lower because fewer subpackages are required, Fig. 8. The 2.5D scaled system interposer is 60% larger than the 3D scaled system, thus increasing EE per die as fewer dies are produced per wafer. When considering 100% die yield, the EE for the interposer in the 2.5D scaled system is 83% higher. Incorporating die yield and GPU package yield increases the EE of the interposer of the 2.5D system by 168% compared to the 3D scaled system.

## IV. Performance

The systems are iso-FLOPs and iso-HBM capacity; however, the 2.5D system provides higher HBM bandwidth due to its larger number of HBM packages. Despite this bandwidth advantage, the 2.5D system is constrained by its network topology: its eight IODs are organized as two switched, fully connected four-IOD groups, limiting connectivity relative to the 3D system, which provides a fully connected topology among its four IODs. This makes the 3D system more effective for workloads with lower compute–communication overlap. We evaluate system performance by partitioning each system around the IODs, see Table III and IV, into logical compute/memory blocks [5] and running training workloads, shown in Table V, using Hespas as the simulation engine [6]. Fig. 9 shows that, although these workloads exhibit high compute–communication overlap, the 2.5D system achieves only a modest 6.8-13.5% speedup.

## V. Conclusions

To reduce the embodied emissions of next generation GPUs used in hyperscale data centers we compare how GPU scaling in 2.5D and 3D impacts EE. The properties and normalized EE for each system are shown in Table I and II. Normalized EE and EE per category are shown in Fig. 5. We find that the 2.5D scaled system increases EE by 141% as compared to 89% for the 3D scaled system while only having modest 6.8-13.5% performance gain. This shows that 3D integration is an attractive option to limit increases in EE by building compact systems that do not significantly impact performance. However, reduced yields from increased HBM stack height and thermal management requirements may become bottlenecks that limit the adoption of 3D technology. Similarly, 2.5D scaled systems also introduce limitations such as communication penalties from longer wire lengths that may degrade performance. Assessing EE in the chip-design phase can help identify carbon-efficient solutions that are necessary to achieve low-carbon AI.

## Acknowledgment

The authors gratefully acknowledge the contributions of the imec SSTS team. This work is funded by the Advanced Research + Invention Agency (ARIA).

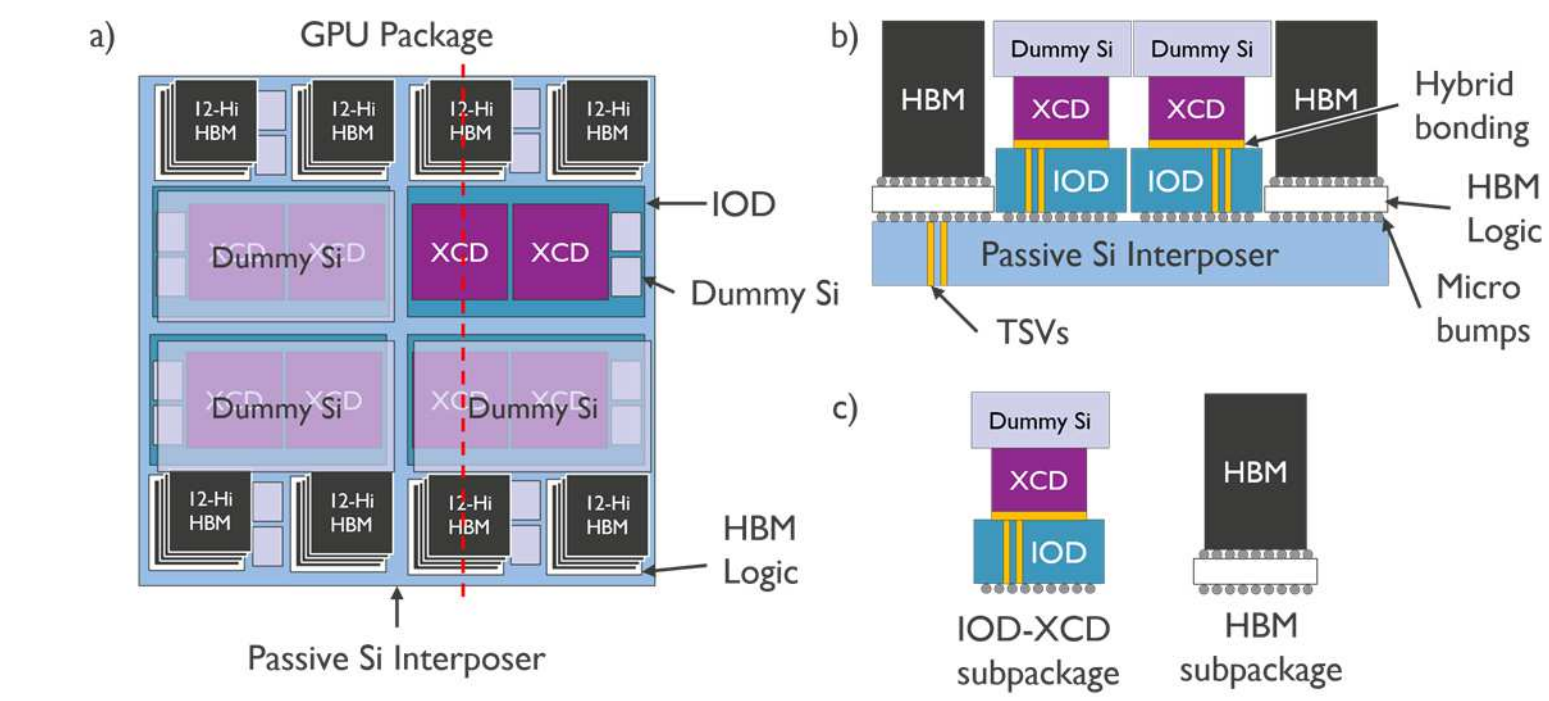

Fig. 1. (a) Top-down view of the GPU package of the baseline system where the top right Dummy Si piece has been removed to show the subcomponents on the IOD, (b) cross-sectional view corresponding to the red dashed line in the top-down view, (c) subpackages that are attached onto the interposer as the final GPU package is assembled.

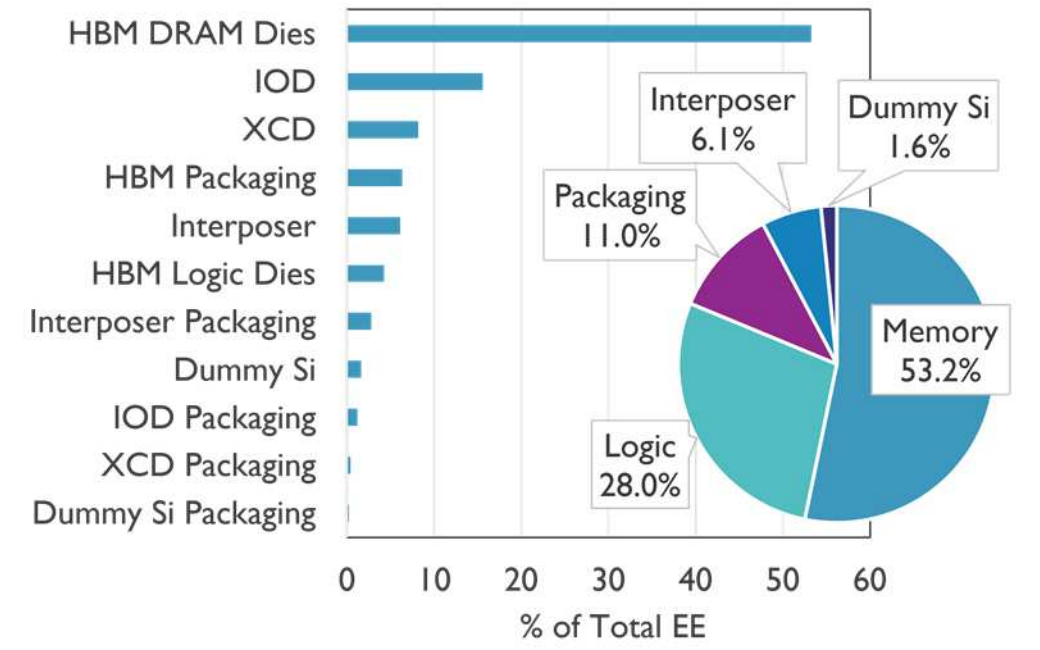

Fig. 2. Breakdown of total EE of the baseline system per component/packaging step and grouped by category. The logic category includes IOD, XCD, and HBM logic dies, the memory category refers to the DRAM dies, and the Dummy Si category refers to the non-functional Si components used for structural stability and thermal dissipation.

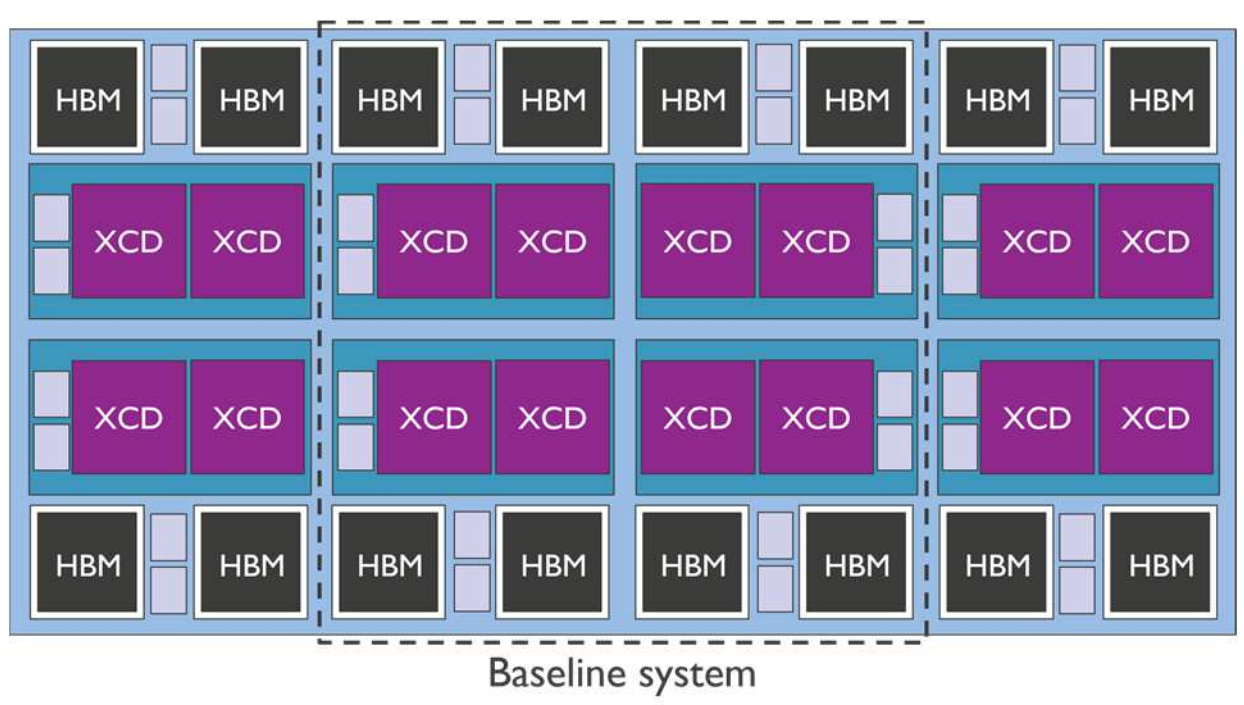

Fig. 3. Schematic of 2.5D scaled system. 16 XCD are hybrid bonded to 8 IOD and are surrounded by 16 stacks of 12-Hi HBM3.

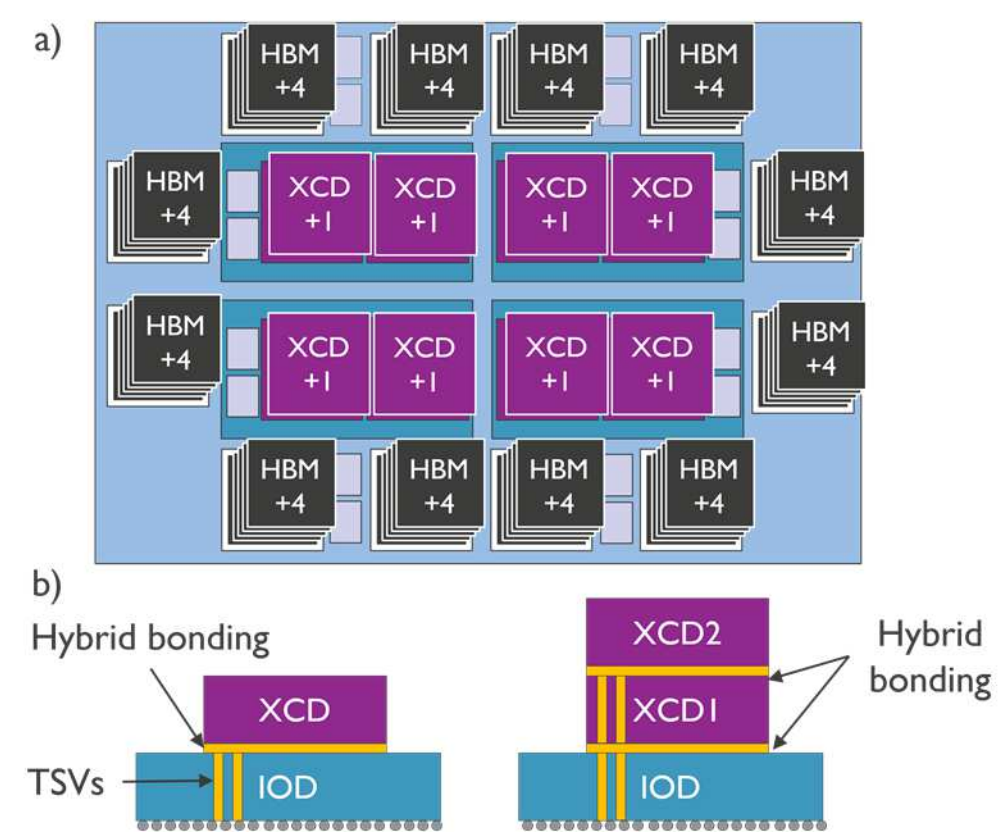

Fig. 4. Schematic of (a) 3D scaled system with 8 additional XCD and 12 stacks of 16-Hi HBM, and (b) IOD-XCD hybrid bonded stack for the baseline and 3D scaled systems.

| | Baseline | 2.5D Scaled | 3D Scaled |
|---|---|---|---|
| Interposer die yield | 64.9% | 43.5% | 58.5% |
| IOD-XCD subpackage yield | 97.0% | 97.0% | 95.1% |
| HBM subpackage yield | 87.8% | 87.8% | 84.3% |
| GPU package assembly yield | 88.6% | 78.6% | 85.1% |

Table. I. Summary of yield values for the baseline, 2.5D scaled, and 3D scaled systems.

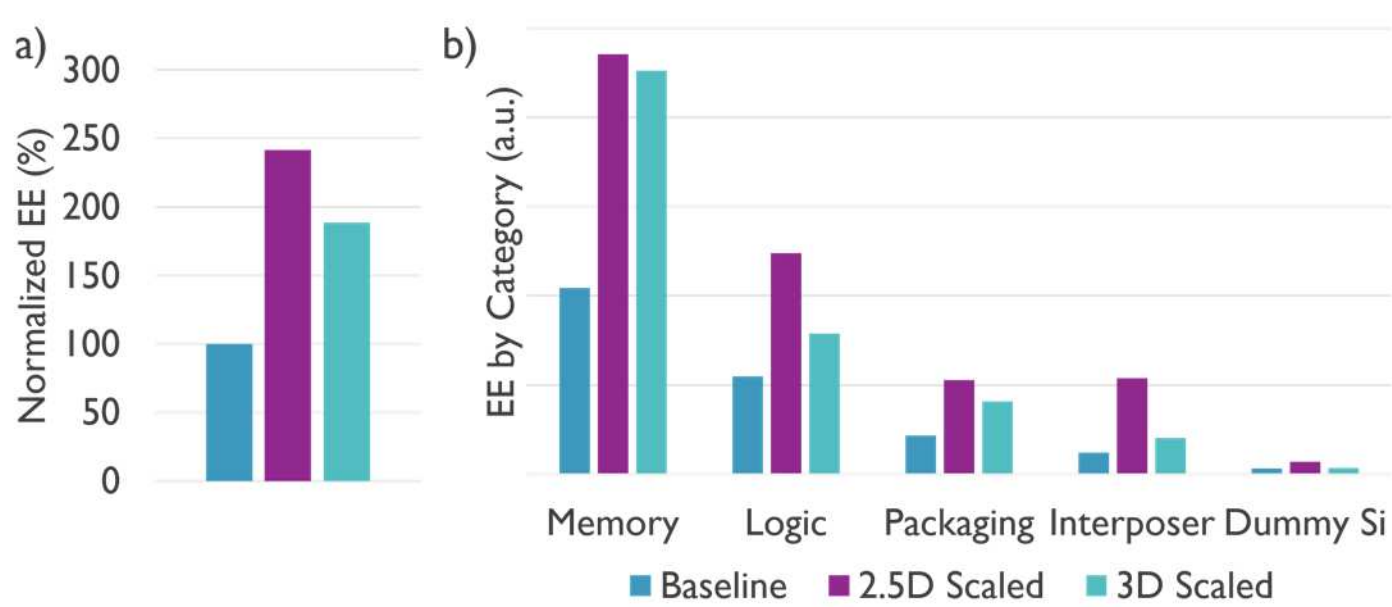

Fig. 5. Comparison of (a) Normalized EE where the 2.5D system has 141% higher EE and the 3D system has 89% higher EE compared to the baseline system. (b) EE by category for the baseline, 2.5D scaled, and 3D scaled systems.

| Parameter | Baseline | 2.5D Scaled | 3D Scaled |
|---|---|---|---|
| #Compute Units | 304 | 608 | 608 |
| Peak Compute FP16 PFLOP/s | 1.3 | 2.6 | 2.6 |
| HBM Capacity (GB) | 192 | 384 | 384 |
| Normalized PFLOP/s | 100% | 200% | 200% |
| Normalized HBM Capacity | 100% | 200% | 200% |
| Normalized EE | 100% | 241% | 189% |

Table II. Summary of compute resources and normalized EE for the baseline system, and the 2.5D and 3D scaled systems. Projected peak FLOP/s is calculated by scaling the published values for the MI300X by the projected number of compute units.

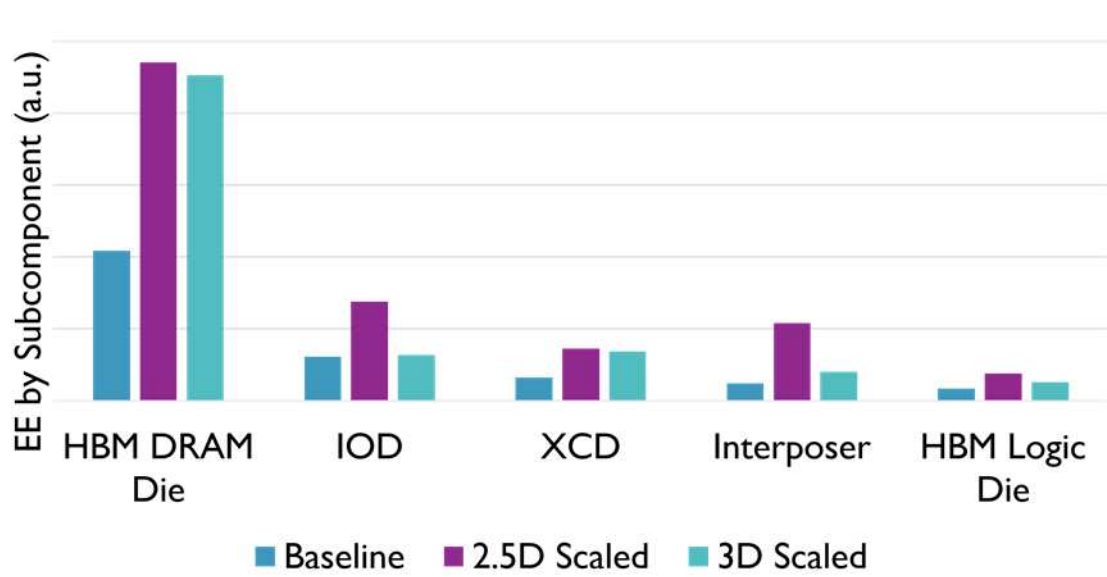

Fig. 6. Comparison of EE for the baseline, 2.5D scaled, and 3D scaled systems for the highest contributing subcomponents.

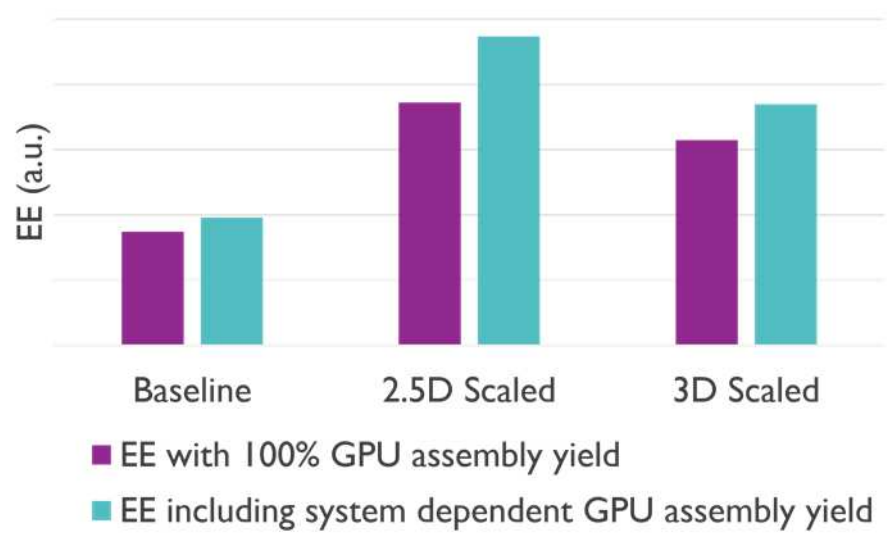

Fig. 7. Impact of GPU assembly yield on EE. GPU assembly yields for the baseline, 2.5D scaled, and 3D scaled systems are 88.6%, 78.6%, and 85.1% respectively.

a)

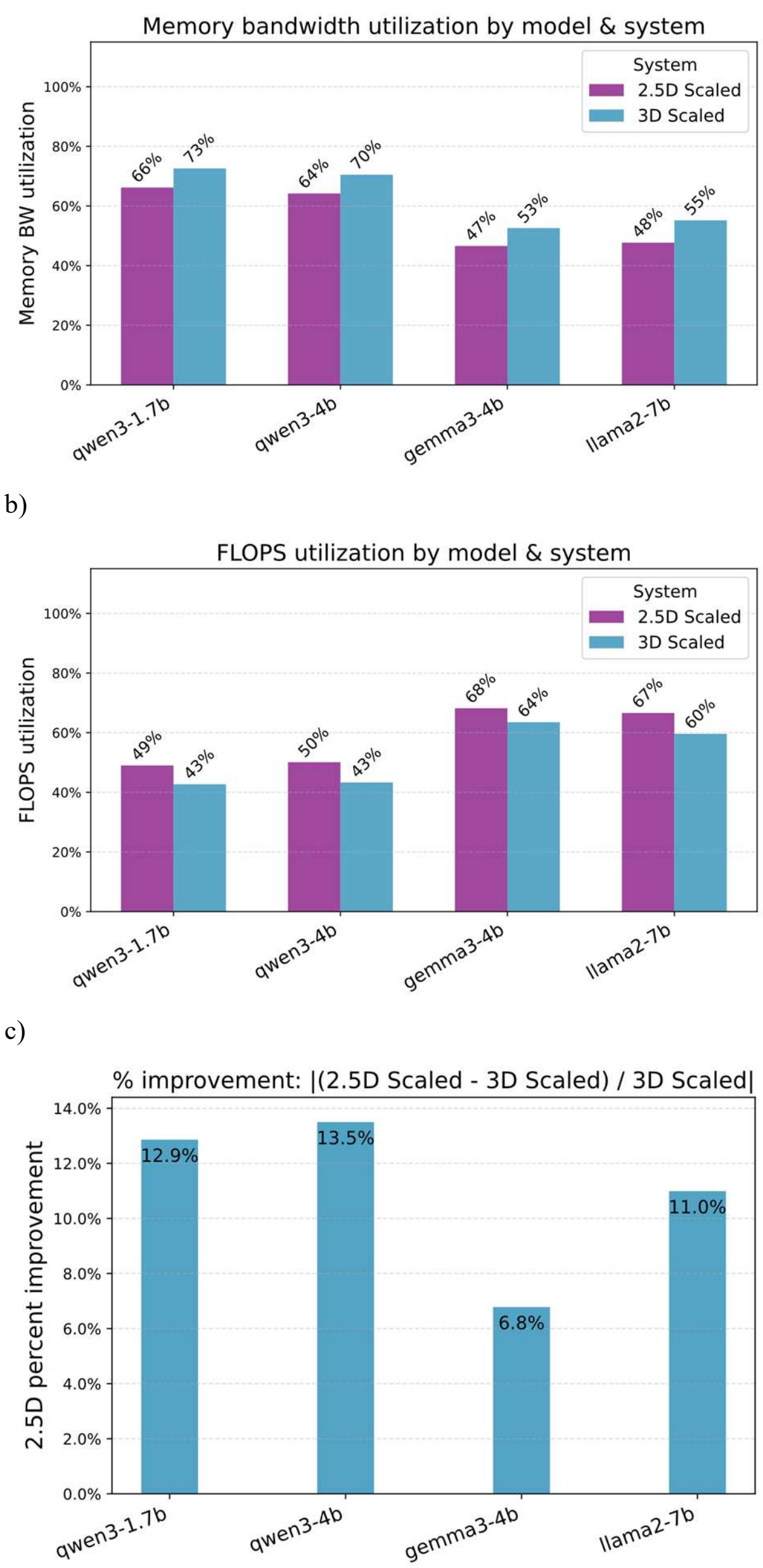

Fig. 9. Comparison of 2.5D and 3D scaled systems for training workloads: (a) memory bandwidth utilization, (b) FLOP/s utilization, and (c) model runtime improvement 2.5D vs 3D. The plots show higher memory pressure for the 3D system and slightly improved compute utilization and lower runtime on the 2.5D system relative to the 3D.

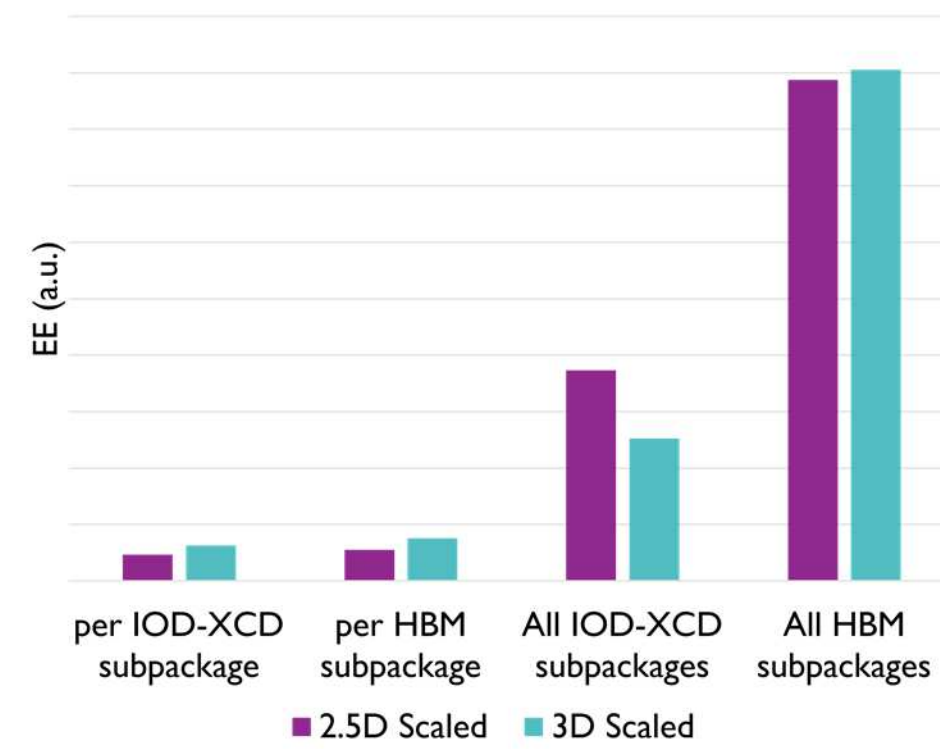

Fig. 8. EE for individual IOD-XCD and HBM subpackages, and total EE for each subpackage for the 2.5D and 3D scaled systems.

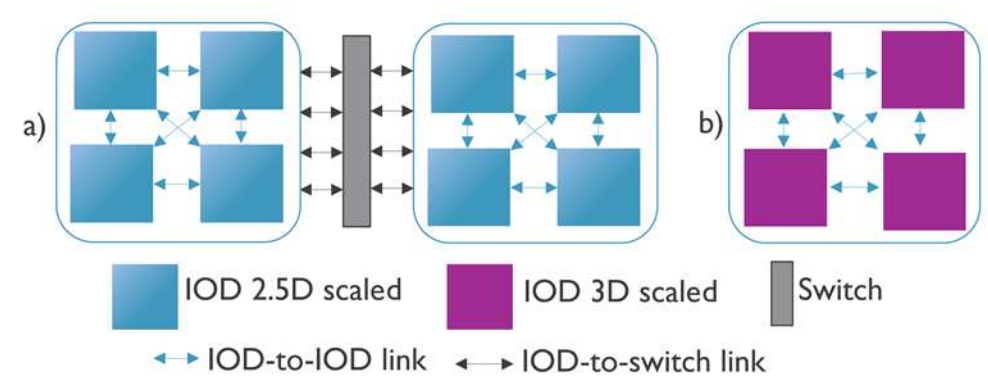

Fig. 10. (a) 2.5D IOD network topology (b) 3D IOD network topology

| System | Perf Model | Peak FLOP/s | Memory BW (GB/s) | HBM Capacity per IOD (GB) | # Logical Partitions |
|---|---|---|---|---|---|
| 2.5D Scaled | Roofline | 325 | 1330 | 48 | 8 |
| 3D Scaled | Roofline | 650 | 1995 | 96 | 4 |

Table. III. Summary of logical compute and memory for 2.5D and 3D scaled systems partitioned around IODs.

| System | Network Topology | NPU count | Link BW (GB/s) | Link Latency (ns) |
|---|---|---|---|---|
| 2.5D Scaled | Fully Connected, Switch | 4, 2 | 4300, 4300 | 200, 200 |
| 3D Scaled | Fully Connected | 4 | 4300 | 200 |

Table. IV. Summary of network configurations for 2.5D and 3D scaled systems partitioned around IODs.

| System | Model | Memory/ IOD (GB) | Data Parallel | Tensor Parallel | Context Parallel |
|---|---|---|---|---|---|
| 2.5D Scaled | qwen3-1.7b | 35.2 | 4 | 1 | 1 |
| | qwen3-4b | 69.4 | 4 | 1 | 1 |
| | gemma3-4b | 62.4 | 4 | 1 | 1 |
| | llama2-7b | 93.2 | 2 | 2 | 1 |
| 3D Scaled | qwen3-1.7b | 21.2 | 8 | 1 | 1 |
| | qwen3-4b | 44 | 8 | 1 | 1 |
| | gemma3-4b | 40.4 | 8 | 1 | 1 |
| | llama2-7b | 48 | 4 | 2 | 1 |

Table. V. Models exported as training workloads in the StableHLO format using the appropriate parallelism scheme to fit the memory capacity of the logical partitions around the IODs.